\documentclass[a4paper,11pt]{article}
\usepackage{jcappub} 
\pdfoutput=1

\usepackage{blindtext}

\usepackage{natbib}
\usepackage{aas_macros}

\usepackage[T1]{fontenc} 
\usepackage{graphicx}
\usepackage{epsfig}
\usepackage{rotate}
\usepackage{amsmath}
\usepackage{amssymb}
\usepackage{amsfonts}
\usepackage{bm,outlines}
\usepackage{diagbox}
\usepackage{slashbox}
\usepackage{verbatim}
\usepackage{tablefootnote}
\usepackage{enumerate}
\usepackage{afterpage}
\usepackage{xcolor}
\usepackage{natbib}
\usepackage{subcaption}
\usepackage{float}

\usepackage[capitalise]{cleveref}
\Crefname{figure}{Fig.}{Figs.}
\Crefname{equation}{Eq.}{Eqs.}

\newcommand{\kpar}{k_{\parallel}}
\newcommand{\kparmin}{k_{\parallel {\rm min}}}

\newcommand{\dif}{\delta^{\rm IF}}
\newcommand{\dsd}{\delta^{\rm SD}}

\newcommand{\bk}{\bm{k}}
\newcommand{\bq}{\bm{q}}

\newcommand{\fnll}{f_{\mathrm{NL}}^{\mathrm{loc}}}

\newcommand{\Mpc}{\ensuremath{\text{$h$/Mpc}}\xspace}

\newcommand*\mean[1]{\overline{#1}}
\newcommand*{\veps} {\varepsilon}

\title{Squeezing information from radio surveys to probe the primordial Universe}

\author[1,2]{Dionysios Karagiannis,}
\author[1,3,4]{Roy Maartens,}
\author[5,6]{Shun Saito,}
\author[7,8,1]{Jos\'e Fonseca,}
\author[9,10,11,1]{Stefano Camera,}
\author[2,1,12]{Chris Clarkson}

\affiliation[1]{Department of Physics \& Astronomy, University of the Western Cape, Cape Town 7535, South Africa}
\affiliation[2]{Department of Physics \& Astronomy, Queen Mary University of London, London E1 4NS, United Kingdom}
\affiliation[3]{Institute of Cosmology \& Gravitation, University of Portsmouth, Portsmouth PO1 3FX, United Kingdom}
\affiliation[4]{National Institute for Theoretical \& Computational Sciences, Cape Town 7535, South Africa}
\affiliation[5]{Institute for Multi-messenger Astrophysics \& Cosmology, Department of Physics, Missouri University of Science \& Technology, 
Rolla MO65409, USA}
\affiliation[6]{Kavli Institute for the Physics \& Mathematics of the Universe, University of Tokyo, Kashiwa, Chiba 227-8583, Japan}
\affiliation[7]{Instituto de Astrof\'isica e Ci\^encias do Espa\c{c}o, Universidade do Porto CAUP, 4150-762 Porto, Portugal}
\affiliation[8]{Departamento de F\'isica e Astronomia, Faculdade de Ci\^{e}ncias, Universidade do Porto, Rua do Campo Alegre 687, PT4169-007 Porto, Portugal}
\affiliation[9]{Dipartimento di Fisica, Universit\`a degli Studi di Torino, 10125 Torino, Italy}
\affiliation[10]{INFN -- Istituto Nazionale di Fisica Nucleare, Sezione di Torino, 10125 Torino, Italy}
\affiliation[11]{INAF -- Istituto Nazionale di Astrofisica, Osservatorio Astrofisico di Torino, 10025 Pino Torinese, Italy}
\affiliation[12]{Department of Mathematics \& Applied Mathematics, University of Cape Town, Cape Town 7701, South Africa}

\emailAdd{dakaragian@gmail.com}

\abstract{
A major goal of cosmology is to understand the nature of the field(s) which drove primordial Inflation. Through future observations, the statistics of large-scale structure will allow us to probe primordial non-Gaussianity of the
curvature perturbation at the end of Inflation. We show how a new correlation statistic can significantly improve these constraints over conventional methods. Next-generation radio telescope arrays are under construction which will map the density field of neutral hydrogen to high redshifts. These telescopes can operate as an interferometer, able to probe small scales, or as a collection of single dishes, combining signals to map the  large scales. We show how to fuse these operating modes in order to measure the squeezed bispectrum with higher precision and greater economy. This leads to constraints on primordial non-Gaussianity that will improve on measurements by Planck, 
and out-perform other surveys such as Euclid. We forecast that $\sigma(\fnll)\sim 3$,  achieved by using a small subset, $\mathcal{O}(10^2-10^3)$, of the total number of accessible triangles. The proposed method identifies a low instrumental noise, systematic-free scale regime, enabling clean squeezed bispectrum measurements. This provides a pristine window into local primordial non-Gaussianity, allowing tight constraints not only on primordial non-Gaussianity, but on any observable that peaks in squeezed configurations.
}

\begin{document}
\maketitle
\flushbottom

\section{Introduction} 

Primordial non-Gaussianity (PNG) is one of the most promising observational avenues to probe the primordial Universe, due to its sensitivity to the physics of these early stages. Measuring PNG could reveal the field content and the amplitude of field interactions \cite{dePutter:2016trg}, as well as verify the quantum nature of the primordial fluctuations \cite{Green_2020} that seed the large-scale structure (LSS) we observe today,  predicted by Inflation. This early Universe information can be extracted by characterizing how much the distribution of the primordial perturbations deviates from Gaussianity.

We propose a novel method that utilises the functionalities of a radio telescope array in  21cm intensity mapping surveys, in order to measure the bispectrum of squeezed triangles -- in an optimal and efficient way, while maximizing the PNG signal and avoiding regions contaminated by systematics. 
The squeezed bispectrum, with  $k_1\approx k_2 \gg\,k_3$, can be used to significantly improve current bounds on  local PNG \cite{Salopek1990,Gangui1993,Verde1999,Komatsu2001}. 
Constraining $\fnll$ provides a unique observational handle to rule out entire classes of inflationary models and enhance our knowledge of the early Universe \cite{Achucarro_2022}.

Currently, the tightest constraints on PNG are from cosmic microwave background (CMB) measurements \cite{Planck_2019PNG}. 
The results imply that the primordial perturbation field is weakly non-Gaussian and hence most of the PNG information is contained in the primordial bispectrum, i.e., the lowest-order deviation from a Gaussian distribution. 
%
Most of the available information on local PNG from the CMB  bispectrum has been extracted. 
LSS surveys have the potential to significantly improve the current PNG bounds  
\cite{Karagiannis2018,Meerburg_2019}, since a 3D dark matter tracer  (e.g. galaxies) has considerably more bispectrum modes than the 2D CMB. 
However, most of these LSS modes are in the nonlinear regime, which is dominated by the gravitational evolution of cosmic structures. 
The subdominant PNG signal must be separated from the late-time non-Gaussian gravitational contributions with significant precision. 
Recent results from optical galaxy surveys \cite{Castorina:2019wmr,Mueller:2021jbt,Cabass:2022wjy,DAmico:2022gki,Cabass:2022ymb,Cagliari_2023}, where an analytical perturbative model has been used (see e.g. \cite{Ivanov:2021kcd}), are far from being competitive with the CMB bounds. 
The main reason is the complexity of accurate analytical models beyond the perturbative regime which is needed to extract the wealth of PNG information from the nonlinear scales. 
A promising alternative is to rely on numerical approaches, like simulation-based \cite{Jung:2022rtn,Coulton:2022qbc,Jung:2022gfa,Coulton:2022rir} or field-level inference \cite{Baumann:2021ykm, Andrews:2022nvv}.
 
In this work, we employ a perturbative analysis at high redshifts, which facilitates access to a large number of bispectrum modes in the perturbative regime. 
By also using high-density samples, we can significantly boost the PNG signal. 
To achieve this, we investigate upcoming high-redshift  21cm intensity mapping surveys \cite{Bharadwaj:2000av,Battye:2004re,Wyithe:2007gz,Chang_2008,Battye2013,Santos:2015gra,Bacon:2018dui, Crichton:2021hlc}, focusing  on the squeezed bispectrum.
Due to its distinct triangle preference, which couples different scales, this shape is mostly orthogonal to the late-time gravitational part of the LSS bispectrum \cite{Schmittfull:2013,Desjacques2016,Karagiannis2018} and therefore relatively easier to separate from the latter.



\section{PNG from 21cm intensity mapping}


In the post-reionization Universe most of the neutral hydrogen (HI) resides inside galaxies, thus detecting its 21cm emission line provides a tracer of the underlying matter distribution. The HI intensity mapping (IM) technique \cite{Chang2007} consists of making brightness temperature maps of the sky at different radio frequencies, by measuring the integrated emission line from many galaxies that reside within each pixel. The combined emission yields a large detectable signal, while probing very large volumes of the Universe from the epoch of reionization up to present times. Such spectroscopic surveys are planned, for example with the Square Kilometre Array Observatory (SKAO) \cite{Bacon:2018dui} and the Hydrogen Intensity and Real-time Analysis eXperiment (HIRAX) \cite{Crichton:2021hlc}, which will cover a wide sky area and deep redshift range ($0<z\lesssim 3$).


 
\begin{figure}[t] 
\centering
\includegraphics[width=0.65\linewidth]{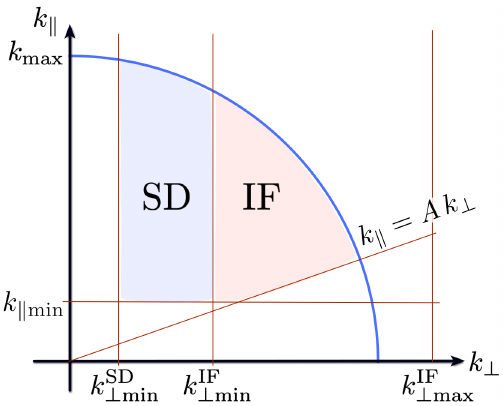}
  \caption{Schematic illustration of the ranges of radial and transverse wavenumbers that the two survey modes are sensitive to. Reproduced from \cite{Bull2015}. © 2015. The American Astronomical Society. All rights reserved. }
\label{fig:plot_krange}
\end{figure}

Radio telescope arrays can measure HI intensity in two distinct ways: in single-dish (SD) mode, by auto-correlating the signal of each dish individually and summing them, or in interferometer mode (IF), where the signals from all dishes are cross-correlated. The first probes the large and linear scales of clustering, while the latter probes the intermediate to the very small and nonlinear scales (large $k$-mode values) with high angular resolution.  
These modes of survey complement each other in Fourier space, see 
\Cref{fig:plot_krange}. Additionally the two functionalities differ in the nature of the instrumental noise, which is the dominant noise contribution \cite{2011ApJ...740L..20G}. Contrary to optical surveys, the shot noise can be negligible, up to small scales. 




In addition to the intrinsic instrumental scale limitations of an HI IM survey, there are known systematics. Astrophysical sources introduce nearly smooth foregrounds which overwhelm the very large-scale fluctuations along the line of sight ($k_\parallel\lesssim 0.01\,\Mpc$) rendering them unusable \cite{Furlanetto:2006jb,Chang2007,Liu2011,Liu2012,Shaw:2013wza,Shaw:2014khi}. 
For an interferometer, an additional  non-smooth foreground component,  the foreground wedge, contaminates some of the modes transverse to the line of sight \cite{Liu2011,Liu2012,Parsons2012,Pober2014,Seo2015,Pober2015,Seo2015}~-- see \Cref{fig:plot_krange}. 

The presence of local PNG induces a large-scale scale-dependent behaviour in the power spectrum of a tracer \cite{Dalal2008,Slosar2008,Matarrese2008,Verde2009,Afshordi2008,Desjacques2010}. Hence, an LSS survey should be able to probe the largest scales of clustering in order to provide competitive power spectrum constraints on $\fnll$. 
For the bispectrum, access to a wide scale range is essential, in order to probe enough squeezed triangles and boost the signal-to-noise ratio. Thus, a single-dish HI IM survey favours a power spectrum analysis 
\cite{Camera2013,Xu2014,Fonseca:2015laa,2017MNRAS.466.2780F, Ballardini:2019wxj}, outperforming an interferometer survey. The reverse holds for the bispectrum $\fnll$ constraints, where surveys designed for IF mode outshine those in SD mode \cite{Karagiannis:2020dpq}. 

Due to the intrinsic instrumental and observational scale restrictions, HI IM surveys have not yet reached their full potential in constraining $\fnll$, despite their optimal specifications (i.e. wide sky area, high redshifts and sample densities). This is even more evident for the bispectrum of an IF survey, where a more futuristic setup, like PUMA \cite{PUMA_surv} or SKA2 LOW \cite{Pourtsidou:2016ctq} is required to significantly improve over current bounds \cite{Karagiannis:2019jjx,Karagiannis:2020dpq} .

\section{The 21cm bimodal bispectrum}

In this work, we propose a novel method to boost the capabilities of  HI IM to tightly constrain the amplitude of local PNG, while excluding the contaminated $k$-modes from the analysis. The key point is that the two functionalities of the radio telescope probe complementary scales (see \Cref{fig:plot_krange}). Combining the data from these two can provide access to a very wide scale range with high resolution: from the large and linear scales, as probed by the SD mode,  to the very small and nonlinear ones, measured by the IF mode. 
This is optimal for a bispectrum analysis and especially for an observable that has a strong signal on squeezed triangles, like local PNG. 

Squeezed configurations correlate different scales ($k_1\sim k_2 \gg k_3$). Thus, cross correlating a density field $\dsd(\bk)$, as probed by an SD mode survey, with two density fields $\dif(\bk)$ probed by a survey in IF mode, we can form an HI IM bispectrum that has access to a very large number of squeezed triangles. We define the bimodal bispectrum, synergising two HI IM surveys in two different operating modes, for an overlapping redshift range and sky area, as
\begin{equation}
    \langle \dif(\bk_1)\,\dif(\bk_2)\,\dsd(\bk_3)\rangle=(2\,\pi)^3 \,\delta_{\rm D}(\bk_{123})\,B^{\rm SD\times IF}(\bk_1,\bk_2,\bk_3)\;,
\end{equation}
where $\bk_{123}\equiv\bk_1+\bk_2+\bk_3$.

The fields $\delta^{\rm SD,IF}$ are fluctuations in the 21cm brightness temperature  and their measurements in Fourier space depend on survey window functions $W^{\rm SD,IF}(\bk)$. Crucially, SD observations observe the actual sky, so that the fluctuation is a convolution in Fourier space, ${\dsd(\bk)=W^{\rm SD}(\bk)\star\delta(\bk)}$, whereas IF mode measures a Fourier transform of the sky, leading to a simple product, ${\dif(\bk)=W^{\rm IF}(\bk)\,\delta(\bk)}$. (We omit the $z$-dependence for brevity.)

The window functions $W^{\rm SD,IF}(\bk)$  
ensure that the measured $k$-modes are well within the intrinsic instrumental limits and away from regions affected by systematics (\Cref{fig:plot_krange}), giving a contamination-free measurement of the synergy bispectrum. These limits include the fundamental limitations of each instrument \cite{Bull2015}, which 
depend on the size of the dishes, as well as the known astrophysical foregrounds and the wedge effect in the case of IF mode. The latter two exclude all scales that satisfy $k_\parallel<\kparmin$ and $k_\parallel<
A(z) \,k_\perp$ \cite{Karagiannis:2020dpq}. 
Foregrounds 
can be recovered with percent precision \cite{Zhu:2016esh,Karacayli:2019iyd,Modi:2019hnu,Makinen:2020gvh,Cunnington:2023jpq}, thus we  consider an optimistic cut-off value, 
$\kparmin=0.005\,\Mpc$.

The estimator for measuring the synergy bispectrum between the SD and IF modes 
is
\begin{equation}\label{eq:estimator}
\hat{B}^{\rm SD\times IF}(\bk_1,\bk_2,\bk_3)=\frac{1}{V_s\,V_{123}}\sum_{\bq_1\in k_1}\sum_{\bq_2\in k_2}\sum_{\bq_3\in k_3}\delta_{\rm K}(\bq_{123})  \,\dif(\bq_1)\,\dif(\bq_2)\,\dsd(\bq_3)\;,
\end{equation}
where $V_s$ is the volume of the survey within each redshift bin. The Kronecker delta ensures the formation of `fundamental triangles' with sides $\bq_i$, satisfying  $\bq_{123}=0$, that fall in the `triangular bin' of width $\Delta k$ defined by the bin centres ($k_1$, $k_2$, $k_3$). The volume in Fourier space of the fundamental triangle bins is, in the thin shell limit $V_{123}= 8\,\pi^2\,k_1\,k_2\,k_3\,(\Delta k) ^3$ \cite{Sefusatti2006}, where the bin size $\Delta k$ is taken to be the fundamental frequency of the survey, $\Delta k=k_{\rm f}=2\pi/V_{\rm s}^{1/3}$. For the extreme cases of flattened ($k_1=k_2+k_3$) and open triangles ($k_1\ne k_2+k_3$) this expression breaks down. Here we consider the analytic corrections for these configurations presented in \cite{Biagetti:2021tua}.

The covariance of the bimodal bispectrum estimator is given by
    ${\sf C}_{ij}\equiv \big\langle \delta \hat{B}_i\,\delta \hat{B}_j\big\rangle={\sf C}_{ij}^{\rm G}+{\sf C}_{ij}^{\rm NG}$\,,
where $\delta \hat{B}_i=\hat{B}_i-\langle\hat{B}_i\rangle$, $i$ and $j$ indices denote the different triangle configurations, i.e.\ $B_i\equiv B(k_1^i,k_2^i,k_3^i)$,  and G, NG are the Gaussian and non-Gaussian contributions. In the thin shell limit ($k\gg \Delta k$), the Gaussian part can be written as
\begin{align}\label{eq:Cov_G}
    {\sf C}_{ij}^{\rm G} =\,\frac{(2\pi)^6}{V_s\,V_{123}}\,s_{123}\,\delta_{ij}\,P^{\rm IF}(\bk_1)\,P^{\rm IF}(\bk_2)\,P^{\rm SD}(\bk_3)\, ,
  \end{align}
where $s_{123}=6,2,1$ for equilateral, isosceles and scalene triangles respectively. The HI IM power spectrum $P^{\rm SD,IF}$ is the tree-level model, including the instrumental noise (see \Cref{sec:appendix} for details).
  

The non-Gaussian part of the covariance is a complicated quantity to calculate analytically. However, in the case of squeezed configurations, like those considered in this work, it can be accurately approximated\footnote{This analytic expression includes the bispectrum-bispectrum and power spectrum-trispectrum contributions. The latter is equivalent to the first in the case of squeezed configurations \cite{Biagetti:2021tua}, thus the factor of 2 in front of \Cref{eq:Cov_NG}. Moreover, the six-point function contribution is negligible for squeezed triangles and for the scales considered here, while the super-sample contribution is subdominant for all scales probed \cite{Barreira:2019icq}.} by \cite{Barreira:2019icq,Biagetti:2021tua,Salvalaggio:2024vmx} 
\begin{align}\label{eq:Cov_NG}
    {\sf C}_{ij}^{\rm NG} =2\,\frac{(2\,\pi)^3}{V_s\,V_{123}^i\,V_{123}^j}
\Big[&\delta_{k_1^ik_1^{\,j}}\,U(k_1^i,k_1^{\,j})\;
 B^{\rm SD\times IF}(\bk_1^{\,j},\bk_2^i,\bk_3^i)\,
 B^{\rm SD\times IF}(\bk_1^i,\bk_2^{\,j},\bk_3^{\,j}) \notag \\&
    +8\,{\rm perm}\Big],   
\end{align}
where $U(k_1^i,k_1^j)=16\pi^3\,k_2^i\,k_3^i\,k_2^{\,j}\,k_3^{\,j}(\Delta k)^5$. For the extreme triangle configurations, we consider the analytic corrections presented in \cite{Biagetti:2021tua}. For $B^{\rm SD\times IF}$, the tree-level model is used and is presented in \Cref{sec:appendix}. A well motivated assumption for the instrumental noise is that it is Gaussian \cite{Bull2015}; in this case the non-Gaussian part of the covariance will not contain instrumental noise contributions.

\section{Results}
We present Fisher forecasts for the HI IM bimodal bispectrum on constraining the amplitude of local PNG. 
In a given redshift bin  the Fisher matrix is
\begin{align}\label{eq:fisherBs}
  F_{\alpha\beta}=\int {\rm d}^3\bk_i \,\frac{\partial B^{\rm SD\times IF}(\bk_i)}{\partial \theta_{\alpha}}\,{\sf C}_{ij}^{-1}\,\frac{\partial B^{\rm SD\times IF}(\bk_j)}{\partial \theta_{\beta}}\,.
  \end{align}
Here $\theta_{\alpha}$ are the parameters of interest, $\bk_i$ is an abbreviation for the three sides of the $i$th triangle, and $\int {\rm d}^3\bk_i\equiv 1/(4\pi)\int_{-1}^{1}d\mu_1\int_0^{2\pi}d\phi\sum_T$, where $\mu_1=\hat{\bk}_1\cdot\hat{\bm z}$ and $\phi$ characterize polar and the azimuthal orientations of a triangle relative to the line-of-sight direction, $\hat{\bm z}$.
The forecasts are from the angle-averaged bispectrum. The sum $\sum_T$ is over all the Fourier space triangles that satisfy $k_{\rm min}\le k_3\le k_2\le k_1 \le k_{\rm max}$. 

We only consider a squeezed subset of these configurations, i.e. those that satisfy $k_1\approx k_2 \ge 3 \, k_{\rm 3}$, since they contain most of the signal for the local PNG. Although the bimodal bispectrum can benefit from all triangle shapes. The minimum value  $k_{\rm min}=k_{\rm f}$,  is the largest scale probed in the redshift bin by the survey, and the maximum value $k_{\rm max}$ corresponds to the smallest scale where the theoretical model is reliable. 
We use a conservative choice $k_{\rm max}(z)=0.75\,k_{\rm NL}(z)$ and an optimistic one $k_{\rm max}(z)=k_{\rm NL}(z)$, where $k_\text{NL}$ is given by the inverse square root of the one-dimensional velocity dispersion \cite{Karagiannis:2019jjx}.
The analysis is thus confined within the perturbative regime, where the tree-level model offers good agreement with the numerical results \cite{Gil-Marin:2014sta,Lazanu2015b,Hashimoto:2017klo,Chan2017,Oddo:2019run}. 
The tree-level description is used to model the expected value of the bimodal estimator \Cref{eq:estimator}, including the observational windows for the two operating modes.
Any $\bk$-mode outside of the window will result in the exclusion of the triangle from the Fisher matrix calculation. 

\begin{figure}[t] 
\resizebox{0.8\linewidth}{!}{\includegraphics{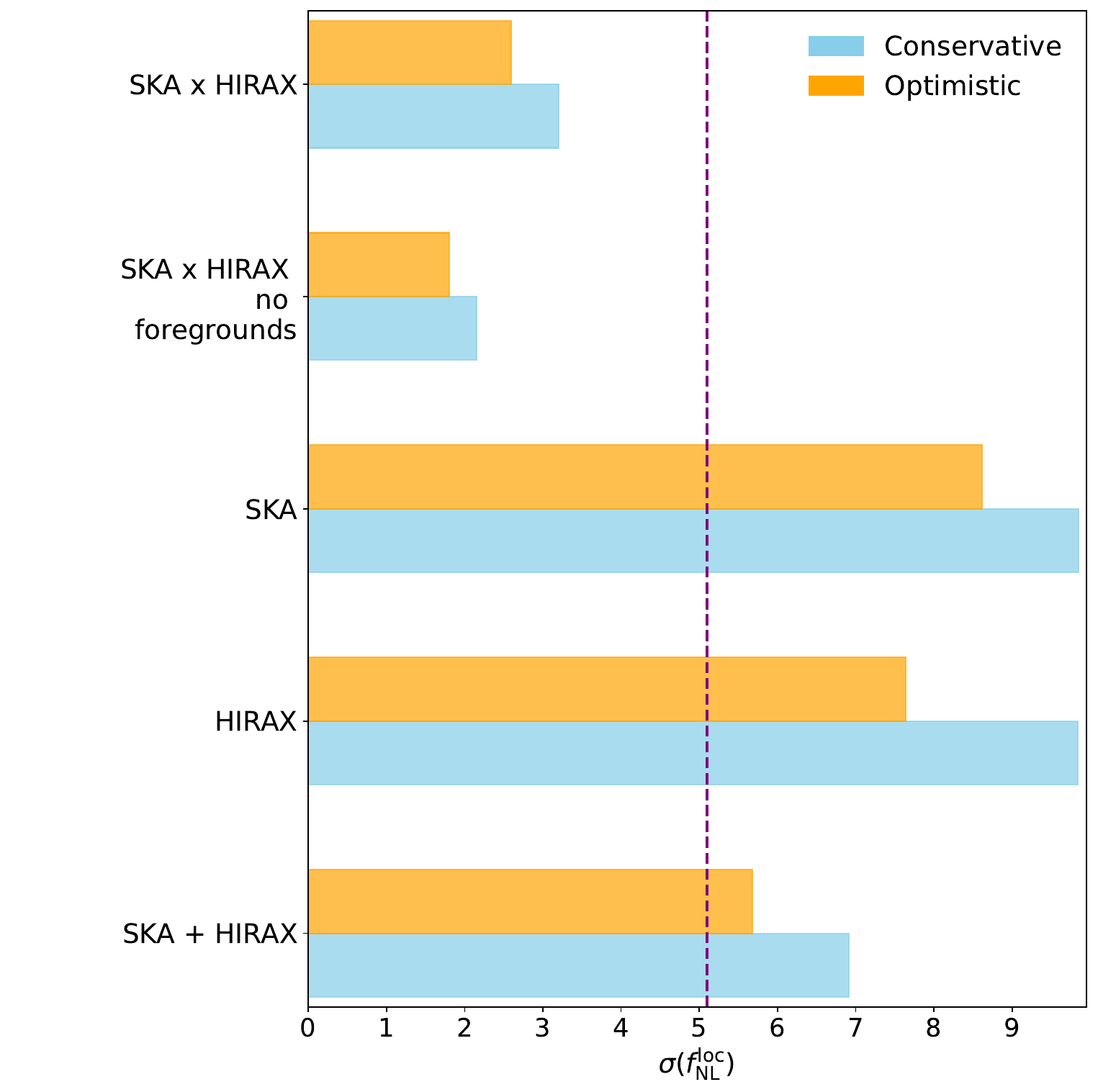}}
  \caption{Marginalized forecasts of $\sigma(\fnll)$ from the bimodal bispectrum (SKA\,$\times$\,HIRAX) with and without foreground cut, for conservative and optimistic $k_{\rm max}$. Forecasts from each survey and from the summed signal without cross correlation (SKA + HIRAX) are also shown. Purple line is the Planck 2018 constraint \cite{Planck_2019PNG}.}
\label{fig:all_surveys_sigma_fnl}
\end{figure}

The forecast error on $\theta_\alpha$ is  given by $\sigma(\theta_\alpha)=(\tilde{F}^{-1})_{\alpha\alpha}^{1/2}$, where $\tilde{F}_{\alpha\beta}=\Sigma_i F_{\alpha\beta}(z_i)$ encompasses the total information from all redshift bins. They are considered independent, with a total number of $N_{\mathrm{bins}}=18$ \footnote{Under the assumption of uncorrelated redshift bins, the final constraints on $\fnll$ show minimal sensitivity to the choice of $\Delta z$. This is supported by additional tests performed with $N_{\mathrm{bins}} = 9$ and $N_{\mathrm{bins}} = 6$, where the resulting changes in the forecasted uncertainties were small ($\sim$1\%).}. All the parameters that control the model are considered free and are entries of the Fisher matrix. These include the $\Lambda$CDM cosmology parameters and the HI IM bias parameters \cite{Karagiannis2018,Karagiannis:2019jjx,Karagiannis:2020dpq,Karagiannis:2022ylq}.

All work on constraining $\fnll$ is affected
by a degeneracy between $\fnll$ and $b_\phi$ \cite{Barreira:2021ueb,Barreira:2022sey}, where $b_{\phi}$ is the local PNG bias parameter \cite{Afshordi2008,McDonald2008}. This compromises the ability of the LSS power spectrum and bispectrum to tightly constrain $\fnll$. PNG bias depends on the halo formation history \cite{Lazeyras:2022koc}, i.e., on properties beyond total mass. 
If it is accurately modeled and constrained, e.g. with hydrodynamical simulations, this degeneracy is broken and the power of the LSS bispectrum to constrain $\fnll$ is completely restored \cite{Fondi:2023egm}. 
Our focus here is on the relative improvement from a new bimodal synergy, compared to the sum of the standard bispectrum information from 2 surveys -- therefore, our results are not affected by the model for $b_\phi$,
and we use the simplest model that breaks the degeneracy \cite{Dalal2008,Matarrese2008,Slosar2008}.

After marginalization over the free parameters of the model \cite{Karagiannis:2020dpq,Karagiannis:2022ylq}, we retrieve the final forecasts on $\fnll$.
The two surveys that we consider are with Band~1 of SKA-MID \cite{Bacon:2018dui}, which is a radio telescope in SD mode and has a redshift range of $z=0.35-3.05$, and the final version of HIRAX \cite{Crichton:2021hlc} in IF mode with $z=0.75-2.55$. 
We consider only the signal from the overlapping range ($z=0.75-2.55$) and sky coverage ($15,000$ $\rm deg^2$).  
For  details of the surveys considered and the specifics of the instrumental noise, see \Cref{sec:appendix} and \cite{Karagiannis:2022ylq}. 

The forecasts on $\fnll$ from the synergy between SKA and HIRAX are presented in \Cref{fig:all_surveys_sigma_fnl}, for the conservative and optimistic $k_{\rm max}$ cases. The results when removing the foreground cut, i.e. when assuming perfect foreground cleaning and signal reconstruction, are shown in order to illustrate the limiting capabilities of the surveys. 
We also present the forecasts from each individual survey, as well as their summed signal (SKA+HIRAX) without cross-correlation, for the same redshift range and squeezed configurations as  the bimodal case. 
The results indicate that \emph{the bimodal bispectrum provides a significant improvement on $\fnll$ constraints by a factor of $\sim4$, compared to those from each individual survey and a factor of $\sim 2-3$ on their summed signal}.

The constraints are competitive with current CMB bounds by a factor of $\sim 2$. The bimodal bispectrum achieves this by only using a small subset, $\mathcal{O}(10^2-10^3)$, of the total number of formed triangles. This means that the forecasts could improve further once all the triangles are considered, even achieving $\sigma(\fnll)\lesssim 1$, as in the case of dedicated surveys with much larger sky area (e.g. SPHEREx \cite{Spherex_2015} or PUMA \cite{Floss:2022wkq}). In addition, by including the single-survey information from the non-overlapping redshifts and sky area,  the constraints would improve even further and utilise to the fullest the available signal in the SKA-HIRAX synergy. 



\begin{figure*}[t]
\centering
  \subfloat{\includegraphics[width=0.45\textwidth]{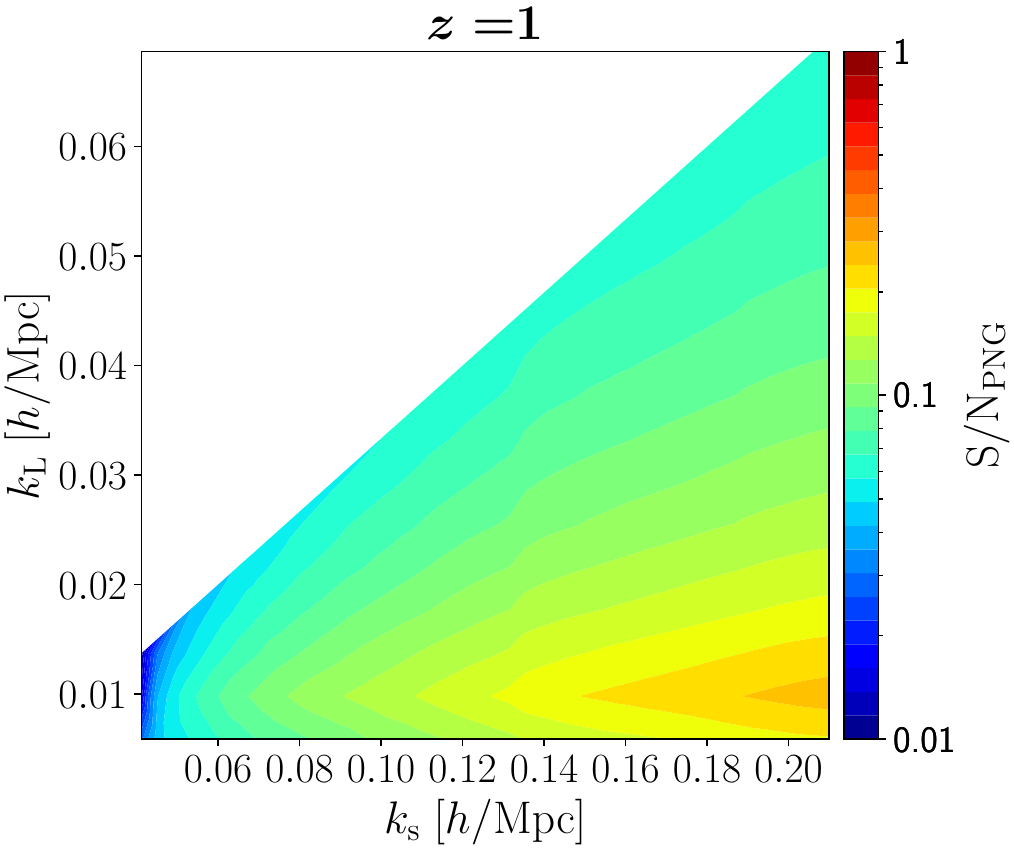}}\hspace{1em}%
  \subfloat{\includegraphics[width=0.45\textwidth]{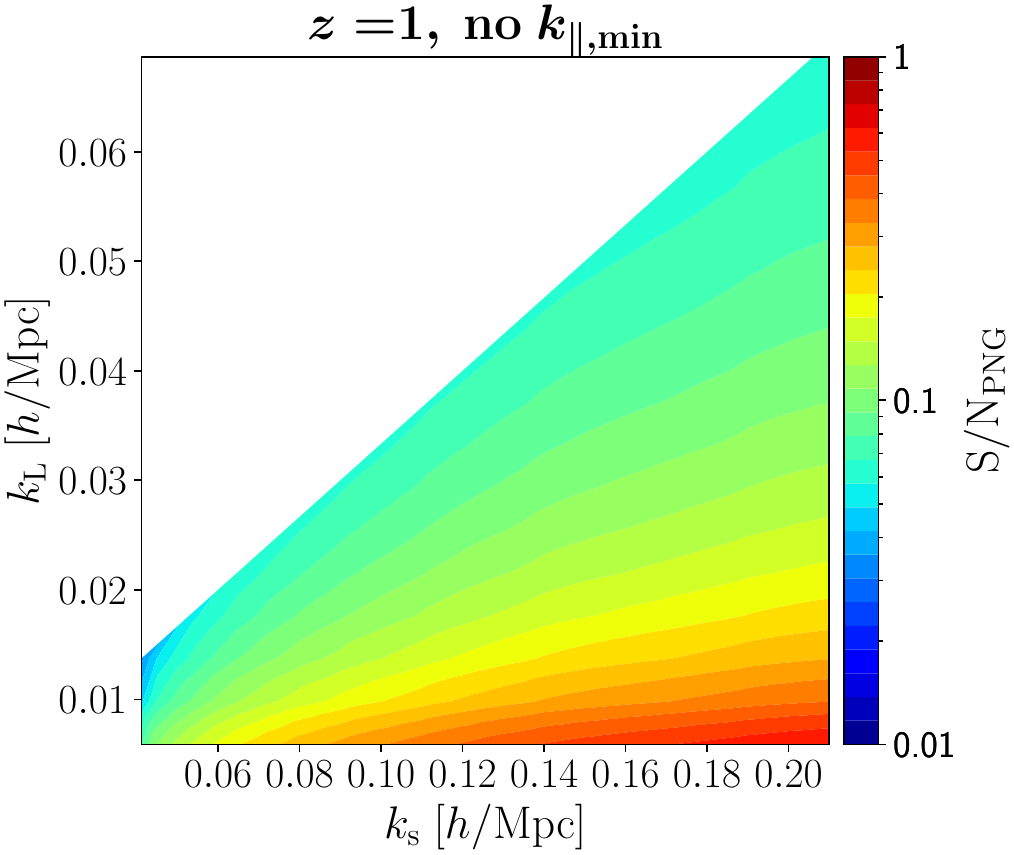}}
  
    \vspace{1.5em} 
  \subfloat{\includegraphics[width=0.45\textwidth]{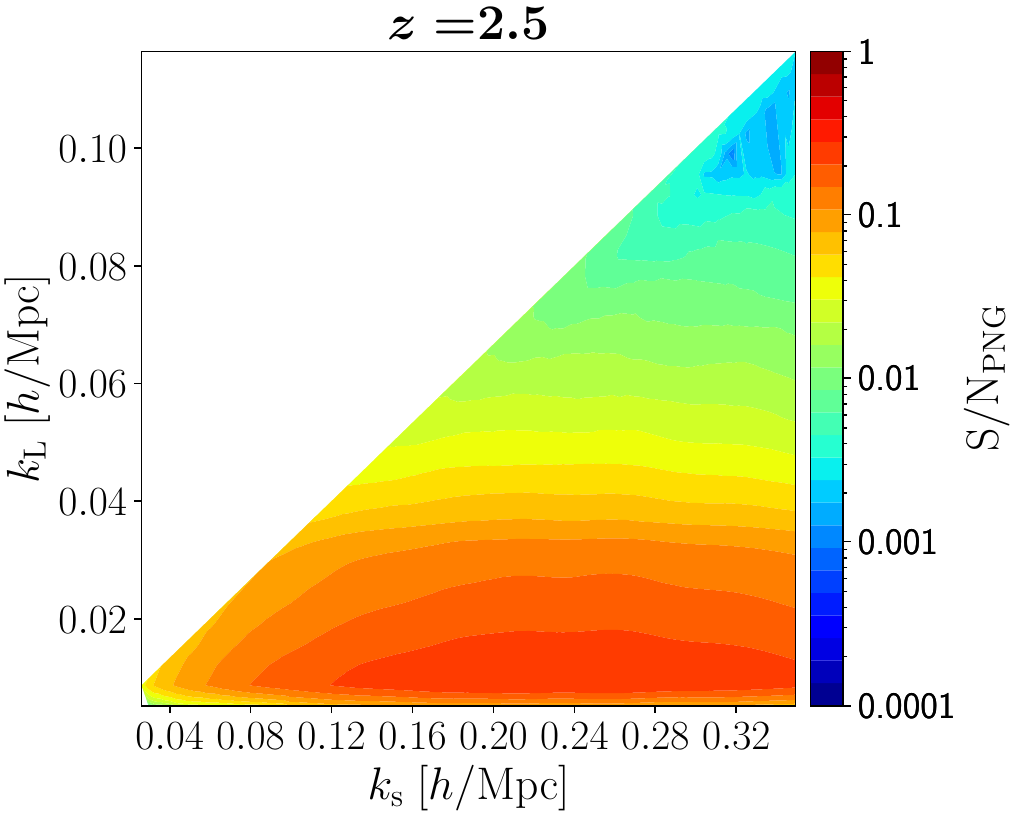}}\hspace{1em}%
  \subfloat{\includegraphics[width=0.45\textwidth]{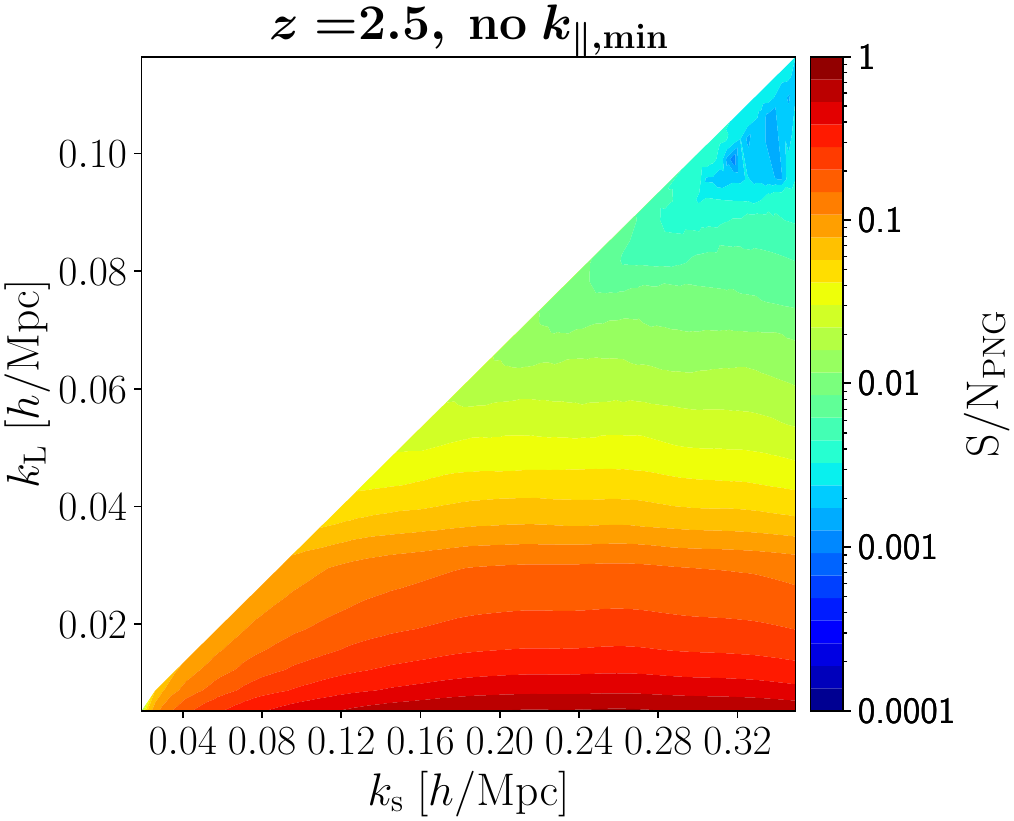}}
  \caption{Normalized non-Gaussian SNR [\Cref{eq:snrng}] as a function of the long- and short-wavelength modes of the squeezed configuration (i.e. $k_1=k_2=k_{\rm S}$ and $k_3=k_{\rm L}$), after integrating over the triangle orientation with the line-of-sight. Upper panels are at redshift $z=1$, lower at $z=2.5$. The left column is the fiducial case, while the right is the case where no radial foregrounds are considered (i.e. no $\kparmin$ cut).
  }
  \label{fig:contour_z}
\end{figure*}


To investigate the effect of the radial foreground cut, we calculate the PNG signal-to-noise ratio (SNR) of the bimodal bispectrum, as
\begin{equation}\label{eq:snrng}
     ({\rm S/N})_{\rm PNG}^2(z)= \int {\rm d}^3\bk \, {B_{\rm PNG}^{\rm SD\times IF}(\bk_i,z)}\,\left({\sf C}^{\rm G}\right)^{-1}_{ij}\, B_{\rm PNG}^{\rm SD\times IF}(\bk_j,z)\;,
\end{equation}
where $B_{\rm PNG}=B(\fnll\ne0)-B(\fnll=0)$. The results are presented in \Cref{fig:contour_z} as a function of the squeezed triangle's sides, i.e. the long-wavelength ($k_3=k_{\rm L}$) and short-wavelength ($k_1=k_2=k_{\rm S}$) modes, after integrating over the triangle orientation angles. Shown at redshifts $z=1$ and $z=2.5$, two scenarios are considered: excluding $k$-modes affected by radial foregrounds ($k_\parallel<\kparmin$, left panels) and fully recovering these modes (right panels).


At low redshifts ($z=1$), recovering the $k$-modes lost to foreground contamination is essential for shifting part of the PNG signal into the linear regime. The dominant contribution to the local PNG signal lies in bispectrum modes where $k_1\sim k_2= k_{\rm S}>k_{\rm max}$, i.e. beyond the range of linear theory. When long-wavelength radial modes are lost to the foregrounds, only a small part of the maximal PNG S/N is accessible. However, if these modes are successfully reconstructed, more squeezed triangles fall within the perturbative regime, making them amenable to analytic modeling and significantly enhancing the contribution of low-redshift data to the PNG constraints. 

In contrast, at high redshifts ($z=2.5$), where the Universe is more linear, the majority of the PNG signal remains within the validity of perturbation theory, even in the presence of a radial foreground cut. As shown in the lower panels of \Cref{fig:contour_z}, these redshifts already contribute significantly to the total PNG constraints. The tree-level model remains reliable, and a large number of squeezed configurations -- particularly those with maximal PNG signal--can still be utilized. While probing smaller scales, i.e. $k_{\rm S}>k_{\rm max}$, would provide access to more configurations, it is not essential for these high-redshift bins to yield strong constraints on $\fnll$.

This explains why, as illustrated in \Cref{fig:all_surveys_sigma_fnl}, reconstructing the lost long-wavelength radial $k$-modes yields a greater improvement in $\fnll$ forecasts than increasing $k_{\rm max}$. Although accessing very large scales is critical for HI intensity mapping, exploring mildly nonlinear regimes offers a complementary strategy to mitigate information loss due to 21cm foregrounds.

To assess the sensitivity of the results to the assumed value of $\kparmin$, we repeat the analysis for $\kparmin = 0.01\;\Mpc$. Increasing $\kparmin$ impacts all surveys and scenarios, with the effect being most significant for the SD mode survey. Specifically, constraints for the SD mode degrade by a factor of $\sim 2.5$ compared to the $\kparmin = 0.005\;\Mpc$ case, while the impact on the IF mode is minimal ($\sim 15\%$). For the bimodal forecasts, the degradation is by a factor of $\sim 2$, and for the summed signal from the SD and IF modes, by a factor of $\sim 1.5$. These results indicate that the advantage of the bimodal approach over the combined signal from the SD and IF surveys remains approximately the same, with a relative improvement by a factor of $\sim 2$.

To evaluate how the ability of the SD survey to probe the large scales influences $\fnll$ constraints, we vary its total integration time and the number of dishes. The results show that even a tenfold reduction in survey capabilities leads to only a a $\sim 10\%$ loss in precision, suggesting that instrumental noise of the SD survey has a limited effect on the constraints. For a more detailed discussion, see \Cref{sec:appendix}.

\section{Conclusions}


We have shown how to combine SD and IF observations of HI IM maps in a novel way to give significantly stronger constraints on $\fnll$ than was previously thought possible. In fact, this method will improve on the Planck results by around a factor of 2, using only a few thousand triangles per redshift bin. The bimodal bispectrum improves constraints relative to the sum of the individual surveys by a factor of $\sim4$,  which is equivalent to a much larger SD or IF array. 

The key aspect of combining SD and IF modes, is that we identified a scale regime where instrumental noise is small and free from other systematics. This enables exceptionally clean measurements of the squeezed bispectrum, containing sufficient squeezed triangle configurations to impose tight constraints on local PNG. This synergy between SD and IF modes is critical for achieving high-precision results on local PNG.

The surveys considered here (SKA and HIRAX) are not the largest possible, as they are intended to serve as test cases to demonstrate the potential of this technique rather than to enable direct comparisons with forecasts from other current or future surveys. Utilizing larger surveys, such as future HI IM projects (e.g., SKA Phase 2, PUMA), could further tighten constraints, surpassing any current forecasts. Indeed, our results already outperform constraints from current galaxy surveys with comparable fields of view to those used here, such as Euclid \cite{Karagiannis2018} and DESI \cite{Rezaie:2023lvi}, with constraints improving substantially as survey parameters are optimized and information from all available triangle configurations is incorporated.

A substantial part of the constraining power comes from the IF mode signal, implying that the results will improve even further if the IF instrumental noise can be reduced. By contrast, the method is not so sensitive to the SD mode noise. Instead, the SD signal is sensitive to the loss of very large scales $\kpar<\kparmin$, so that further improvement requires advances in foreground cleaning and long-wavelength $\kpar$-mode reconstruction. The sensitivity of the results to the assumed value of $\kparmin$ is also examined. While all cases are impacted, the advantage of the bimodal approach over the combined signal from the SD and IF modes remains largely unchanged, highlighting the effectiveness of the proposed approach in constraining local PNG from HI IM surveys.

Long-wavelength modes provide a set of very squeezed triangles that carry a significant amount of signal on local PNG. On the other hand, increasing the number of squeezed triangles by changing $k_{\rm max}$ provides a much less pronounced gain in the $\fnll$ constraints ($\sim 20\%$). The constraints with an optimistic $k_{\rm max}$ are very close ($\sim 15\%$ difference) to those provided by the conservative case, in which no radial foregrounds are considered. This indicates that the local PNG signal lost to foregrounds can be partially recovered, by venturing further into the intermediate regime. Smaller scales have a significant amount of information on PNG, which requires alternative methods to the analytical modelling, e.g. simulation-based inference \cite{Jung:2022rtn,Coulton:2022qbc,Jung:2022gfa,Coulton:2022rir}. We leave this for future work.

The bimodal bispectrum cross-correlates SD large scales with smaller-scale IF modes. We do not expect foregrounds or instrumental systematics to show correlations between long- and short-wavelength $k$-modes, another advantage of the bimodal synergy. The advantages of the bimodal bispectrum are not limited to local PNG studies. Any observable that has signal peaking on squeezed triangles can benefit from the proposed synergy. Moreover, including all triangle configurations into the bimodal analysis, still maintains a significant gain in the constraining power (see \Cref{sec:appendix}).

In the case of bispectrum probes of optical surveys, the data are observed in real space, thus a Fourier transformation takes place from the observed three-point correlation function, including the window functions for each side of the triangle. This introduces complex convolutions between the windows and significantly hinders a bispectrum analysis. In the case of the bimodal synergy bispectrum, the IF mode measures a Fourier transformation of the sky, while the SD mode the actual sky, including the observational windows. This means that the window convolutions will be much more under control, offering an additional benefit to the proposed approach.

\acknowledgments
 DK is supported by the Science and Technology Facilities Council (UK) grant number ST/X000931/1.  DK, RM and SS acknowledge support for this work from the University of Missouri South African Education Program. RM is supported by the South African Radio Observatory and National Research Foundation (grant no.\ 75415). 
SS acknowledges support for this work from NSF-2219212. 
SS is supported in part by World Premier International Research Center
Initiative, MEXT, Japan.
SC acknowledges support from the Italian Ministry of University and Research, PRIN 2022 `EXSKALIBUR – Euclid-Cross-SKA: Likelihood Inference Building for Universe Research', from the Italian Ministry of Foreign Affairs and International
Cooperation (grant no.\ ZA23GR03), and from the European Union -- Next Generation EU. JF acknowledges support of Funda\c{c}\~{a}o para a Ci\^{e}ncia e a Tecnologia through the Investigador FCT Contract No. 2020.02633.CEECIND/CP1631/CT0002, the FCT project PTDC/FIS-AST/0054/2021, and the research grants UIDB/04434/2020 and UIDP/04434/2020.

\clearpage

\appendix
\section{HI IM model and forecast details} \label{sec:appendix}

The surveys considered here are the SKAO\footnote{\url{www.skatelescope.org}} \cite{Bacon:2018dui}, operating in single-dish mode, and HIRAX\footnote{\url{hirax.ukzn.ac.za}} \cite{Crichton:2021hlc}, operating in interferometer mode. The survey specifications are presented in \Cref{table:survey_specs}. 

\begin{table}[t]
 \centering
 \begin{tabular}{l|c|c}
  & SKAO (Band 1) & HIRAX\\ \hline\hline
   redshift   &  $0.775-2.55^{a}$ & $0.775-2.55$  \\
 $N_{\rm dish}$ & $197$ & $1,024$\\
 $D_{\rm dish}$ [m] & $15$ & $6$ \\
 $S_{\rm area}$ [$\rm{deg}^2$] & $15,000^{b}$ & $15,000$ \\
 $t_{\rm survey}$ [hrs] & $10,000$ & $17,500$
 \end{tabular}
 \caption{The survey specifications for the two HI intensity mapping (IM) surveys considered in this work. Notes: (a) The full redshift range covered by SKAO Band 1 is $z = 0.35 - 3.05$ (see Ref.~\cite{Bacon:2018dui} for details). However, to quantify only the gain of the bi-modal approach, i.e. , we restrict our analysis to the overlapping redshift range, $z = 0.775 - 2.55$, which matches that of HIRAX. (b) While SKAO offers a total sky coverage of $20,000$ deg$^2$, we limit our the analysis to the overlapping sky area of $15,000$ deg$^2$, again to ensure a consistent comparison. }
 \label{table:survey_specs}
\end{table}

The tree-level HI IM power spectrum and  bispectrum model in redshift space for  operational modes SD, IF are given by :
\begin{align}
     P^{\rm SD,IF}(\bk,z)&=T_b(z)^2\Big[D_\text{FOG}^P(\bk,z)Z_1(\bk,z)^2P_m^{\rm L}(k,z)+P_{\veps}(z)\Big]+P_{\rm N}^{\rm SD,IF}(\bk,z) \label{eq:Pgs},\\ 
   B^{\rm SD \times IF}(\bk_1,\bk_2,\bk_3,z)&= T_b(z)^3\, \Bigg[D_\text{FOG}^B(\bk_1,\bk_2,\bk_3,z)\nonumber \\ 
   &~~\times\bigg[Z_1(\bk_1,z)Z_1(\bk_2,z)Z_1(\bk_3,z)B_{\rm p}(k_1,k_2,k_3,z) \nonumber \\ 
   &~~+\Big[2Z_1(\bk_1,z)Z_1(\bk_2,z)Z_2(\bk_1,\bk_2,z)P_m^{\rm L}(k_1,z)P_m^{\rm L}(k_2,z)+2~ \text{perm}\Big]\bigg] \nonumber \\
   &~~+2P_{\veps\veps_{\delta}}(z)\Big[Z_1(\bk_1,z)P_m^{\rm L}(k_1,z)+2~ \text{perm}\Big]+B_{\veps}(z)\Bigg]. \label{eq:Bgs} 
  \end{align}
In the equations above we omit the window functions for brevity. Here $P_m^{\rm L}$ is the linear matter power spectrum, as given by CAMB \cite{CAMB}, while $B_{\rm p}$ is the primordial bispectrum for local PNG.  The instrumental noise $P_{\rm N}^{\rm SD,IF}$ is different for each operational mode, with different scale behaviour, and depends on the HI IM survey specifications: see \cite{Bull2015,Santos:2015gra,Karagiannis:2020dpq} for the expressions and details. The background temperature is given by $T_b=188\,\Omega_{\rm HI} (z)h(1+z)^{2}H_0/H(z)~\mu$K, while the HI density evolution is modelled as $\Omega_{\rm HI}(z)=4\times 10^{-4}(1+z)^{0.6}$ \cite{Battye2013}. The stochastic contributions are given by $P_{\veps}(z)=1/\mean{n}(z)$, $P_{\veps\veps_{\delta}}(z)=b_1(z)/\mean{n}(z)$ and $B_{\veps}(z)=1/\mean{n}^2(z)$ \cite{Schmidt2015,Desjacques2016}, where $\mean{n}$ is the effective number density of HI IM \cite{Castorina2016}. These shot-noise terms can be neglected for HI IM (in this work they are considered) since they are small for the scales we consider \cite{Santos:2015gra,Bull2015}. Thus, the instrumental noise is the dominant noise contribution for an HI IM. The details and expressions used to calculate this term for the two instrumental modes can be found e.g. in \cite{Karagiannis:2022ylq}. The effect of the velocity dispersion, the fingers-of-god effect (FoG), is described by a phenomenological by a damping factor \cite{Jackson1972,Peacock1994,Ballinger1996} and given by $D_\text{FOG}^P(\bk)=\exp\big[-\big(k\mu\sigma_P\big)^2\big]$ and $D_\text{FOG}^B(\bk_1,\bk_2,\bk_3)=\exp\big[-\big(k_1^2\mu_1^2+k_2^2\mu_2^2+k_3^2\mu_3^2\big)\sigma_B^2\big]$ for the power spectrum and bispectrum respectively. The damping amplitudes $\sigma_P$ and $\sigma_B$ are given by the linear  velocity dispersion $\sigma_\upsilon$.

The redshift kernels up to second order, which include local PNG, are \cite{Baldauf2011,Tellarini2016,Karagiannis2018}:
\begin{align}  Z_1(\bk_i,z)&=b_1(z)+f(z)\mu_i^2+\frac{\fnll b_\phi(z)}{M(k_i,z)}, \label{eq:Z1}\\
Z_2(\bk_i,\bk_j,z)&=b_1(z)F_2(\bk_i,\bk_j)+f(z)\mu_{ij}^2G_2(\bk_i,\bk_j)+\frac{b_2(z)}{2} +\frac{b_{s^2}(z)}{2}S_2(\bk_i,\bk_j)   \nonumber \\ 
   &~~+\frac{f(z)\mu_{ij}k_{ij}}{2}\left[\frac{\mu_i}{k_i}Z_1(\bk_j,z)+\frac{\mu_j}{k_j}Z_1(\bk_i,z)\right]
   \nonumber \\ 
   &~~+\fnll\Bigg[\frac{b_{\phi\delta}(z)-b_{\phi}(z)N_2(\bk_j,\bk_i)}{2M(k_i,z)}+\frac{b_{\phi\delta}(z)-b_{\phi}(z)N_2(\bk_i,\bk_j)}{2M(k_j,z)}\Bigg]. \label{eq:Z2}
  \end{align}
where $f(z)$ is the linear growth rate, $\mu_i=\hat\bk_i\cdot\hat{\bm{z}}$, with $\hat{\bm z}$ being the line-of-sight vector, $\mu_{ij}=(\mu_ik_i+\mu_jk_j)/k_{ij}$ and $k_{ij}^2=(\bk_i+\bk_j)^2$. 
The function $M(k,z)=2c^2D(z)\,T(k)\, k^2/(3\Omega_m H_0^2\, {g_{\rm dec}})$, contains the linear growth factor $D(z)$ and  matter transfer function $T(k)$, and  $g_{\rm dec}$ is the Bardeen potential growth factor at decoupling (ensuring that $\fnll$ is in the CMB convention \cite{Camera:2014bwa}).
The kernels $F_2(\bk_i,\bk_j)$ and $G_2(\bk_i,\bk_j)$ are the second-order symmetric SPT kernels \cite{Bernardeau2002}, while $S_2(\bk_1,\bk_2) = (\hat\bk_1\cdot\hat\bk_2)^2-1/3$ is the  tidal kernel \cite{McDonald2009,Baldauf2012}. The kernel $N_2(\bk_1,\bk_2) = (\bk_1\cdot\bk_2)k_1^2$  encodes the coupling of the PNG potential to the Eulerian-to-Lagrangian displacement field  \cite{Giannantonio2010,Baldauf2011}. The bias parameters $b_1$ and $b_2$ are the HI linear and quadratic bias parameters, $b_{s^2}$ is the tidal bias coefficient, while $b_\phi$ and $b_{\phi\delta}$ are the local PNG bias parameters.

\begin{table}[t]
 \centering
 \begin{tabular}{c c c c c c}
  $\Omega_{\rm b}$ & $\Omega_{\rm c}$ & $h$ & $A_{\rm s}$ & $n_{\rm s}$ & $\fnll$  \\ \hline\hline
    $0.0493$ & $0.2644$ & $0.6736$ & $2.1 \times 10^{-9}$ & $0.96488$ & $0$
 \end{tabular}
 \caption{The fiducial values used for the flat $\Lambda$CDM cosmology, as measured by Planck \cite{Planck2018_cosmo}. }
 \label{table:cosmo_fid}
\end{table}

The full set of parameters consists of 5 cosmological parameters, the primordial PNG amplitude and 11 further parameters in each redshift bin $i$:
 \begin{align}\label{eq:params}
 \bm{\theta}(z_i)=&\Big\{\Omega_{\rm b},\Omega_{\rm c},h,A_{\rm s},n_{\rm s},\fnll; \\\nonumber
  &~~~    D_A(z_i),H(z_i),f(z_i),b_1(z_i),b_2(z_i),b_{s^2}(z_i),\sigma_P(z_i),\sigma_B(z_i),P_\veps(z_i),P_{\veps\veps_{\delta}}(z_i),B_{\veps}(z_i) \Big\}.
  \end{align}
For the 5 cosmological parameters we add prior information from the observation of CMB performed by Planck \cite{Planck2018_cosmo}. We assume redshift bins are independent, so that e.g.  $\partial P(\bk,z_i)/\partial \theta_\alpha(z_j)=0$ for $i\neq j$. All the above parameters are considered unknown and are marginalized over in the Fisher matrix formalism. Note that we also consider the anisotropies induced in the observed galaxy clustering by the Alcock-Paczynski effect \cite{Alcock1979}, which occurs when the fiducial cosmology, used to convert the observed angular coordinates and redshifts to physical distances, differs from the true one \cite{Seo2003,Song2015}. The fiducial values used for the cosmological parameters are those measured by Planck \cite{Planck2018_cosmo} and are presented for convenience in \Cref{table:cosmo_fid} . The redshift-dependent expressions for the HI bias parameters are provided in \cite{Karagiannis:2022ylq}. The PNG bias parameters, $b_{\phi}(b_1)$ and $b_{\phi\delta}(b_1)$, are considered a function of the linear bias, i.e. $b_{\phi}(b_1)$ and $b_{\phi\delta}(b_1)$, using the `universality relation' \cite{Desjacques2016}, where the expressions are presented in e.g. \cite{Karagiannis:2020dpq}. These PNG bias parameters are considered fixed throughout the analysis.

\begin{figure*}[t]
\centering
  \subfloat{\includegraphics[width=0.45\linewidth]{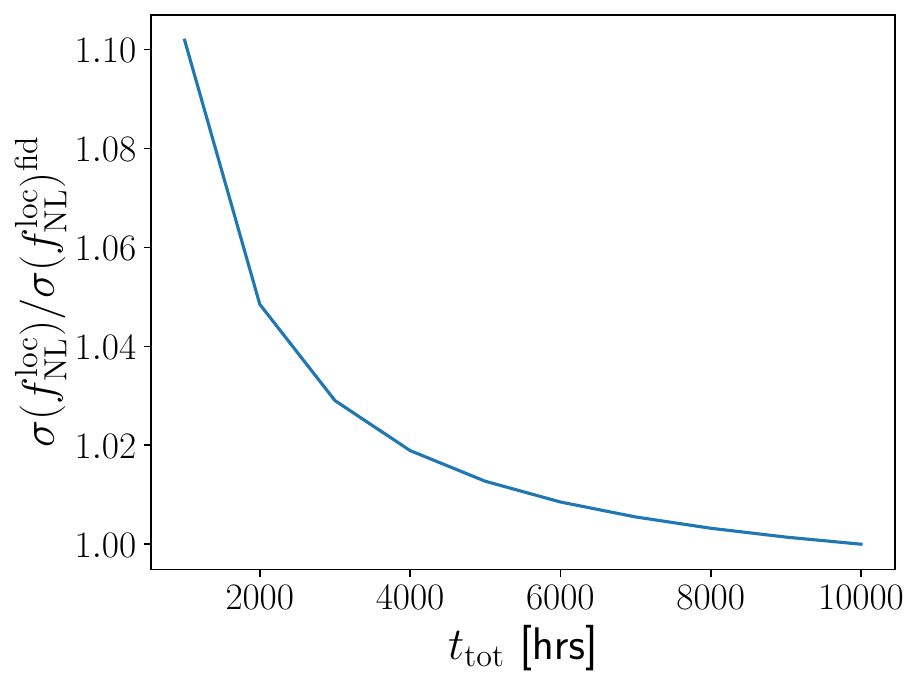}}\hspace{1em}%
  \subfloat{\includegraphics[width=0.45\linewidth]{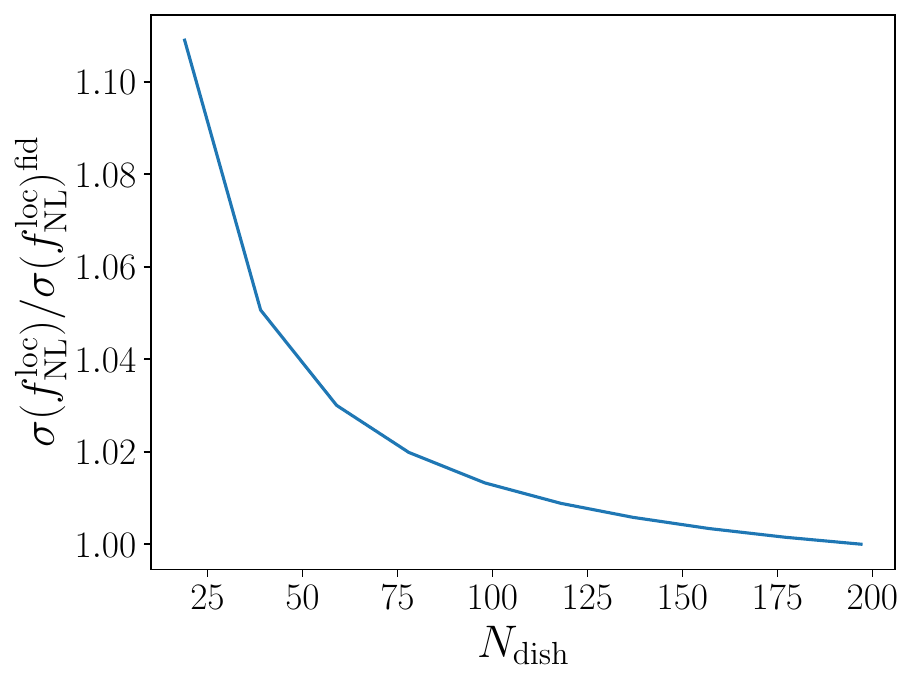}}
  \caption{Marginalized 1$\sigma$ forecasts on $\fnll$, normalized to the fiducial constraints (i.e. SKA\,$\times$\,HIRAX in \Cref{fig:all_surveys_sigma_fnl} ) as a function of the time of the single-dish survey $t_{\rm tot}$ (left panel) and the number of dishes $N_{\rm dish}$ (right panel).
  }
  \label{fig:vary_SD_specs}
\end{figure*}

As emphasized in the main text, accessing sufficiently long-wavelength $k$-modes with the SD survey is important for the proposed method to deliver competitive $\fnll$ constraints, irrespective of whether radial foregrounds are mitigated. This depends on how effectively the SD survey can probe large scales. To investigate this further, we generate $\fnll$ forecasts  using the methodology described in the main text, varying key instrumental parameters that influence the ability of the SD survey (i.e., SKA) to probe large scales. Specifically, we consider the total integration time ($t_{\rm tot}$) and the number of dishes ($N_{\rm dish}$), both of which impact the instrumental noise. Increasing either parameter leads to reduced instrumental noise levels and, consequently, improved precision in $\fnll$ measurements.

\begin{figure}[t] 
\resizebox{0.8\linewidth}{!}{\includegraphics{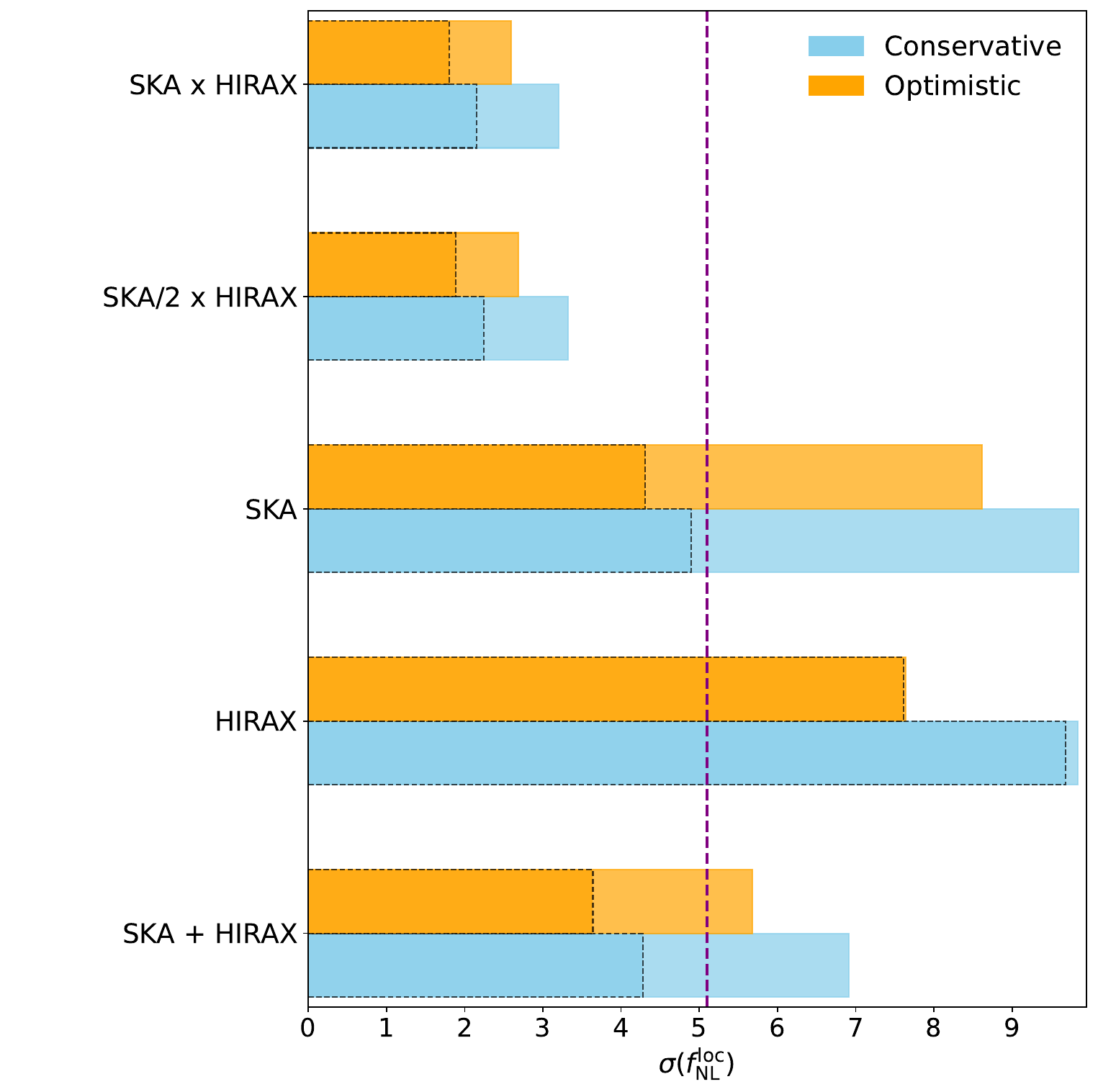}}
  \caption{Marginalized 1$\sigma$ forecasts on $\fnll$ from the bimodal bispectrum  (SKA\,$\times$\,HIRAX), in the case of the conservative and optimistic $k_{\rm max}$. A bimodal case with the capability of the SD survey (SKA) significantly diminished (the number of dishes and observational time are reduced to half), is also presented (SKA/2\,$\times$\,HIRAX). Forecasts from each individual survey, as well as their summed signal (SKA + HIRAX) are  presented for comparison. Planck 2018 constraints \cite{Planck_2019PNG} are indicated with a purple line. The dashed line bars correspond to the idealized case, where the radial signal lost to foregrounds is recovered (i.e. no $\kparmin$ cut is applied). This represents the maximal constraints from the proposed bimodal method. }
\label{fig:all_surveys_sigma_fnl2}
\end{figure}

The forecast $\sigma(\fnll)$, normalized to the fiducial SD survey specifications used to derive the main results (\Cref{fig:all_surveys_sigma_fnl}), is shown in \Cref{fig:vary_SD_specs} as a function of $t_{\rm tot}$ (left panel) and $N_{\rm dish}$ (right panel). Decreasing by an order of magnitude the capabilities of the SD mode survey (e.g. considering only 20 dishes out of 200 or 1000 hrs instead of 10000 hrs), degrades the $\fnll$ constraint by only $\sim10\%$. This indicates that the effect of the instrumental noise of the SD mode survey is small, since it is still able to efficiently probe the large scales. 

We use as an example a variation of the SD survey, `SKA/2', where the integrated time $t_{\rm tot}$ and the number of dishes $N_{\rm dish}$ are reduced by half. The bimodal bispectrum $\fnll$ forecasts are presented in \Cref{fig:all_surveys_sigma_fnl2} (SKA/2\,$\times$\,HIRAX) by following the process described in the main text. Moreover, the idealized case where the radial $k$-modes lost to foregrounds are reconstructed  (i.e. no $\kparmin$ cut), is also shown in dashed lines. The results show that reducing the capabilities of the SD survey by half, has minimal effect on the bimodal constraints on $\fnll$. This is true for both values of $k_{\rm max}$ (`conservative' and `Optimistic') and for the cases where  the modes $k_\parallel<\kparmin$ are included (dashed lines) and excluded (solid lines) from the analysis. The tightest constraint, $\sigma(\fnll)\sim 2$, corresponds to the optimistic $k_{\rm max}$ and the case with no $\kparmin$ cut.

\begin{figure*}[t] 
\centering
\resizebox{0.8\linewidth}{!}{\includegraphics{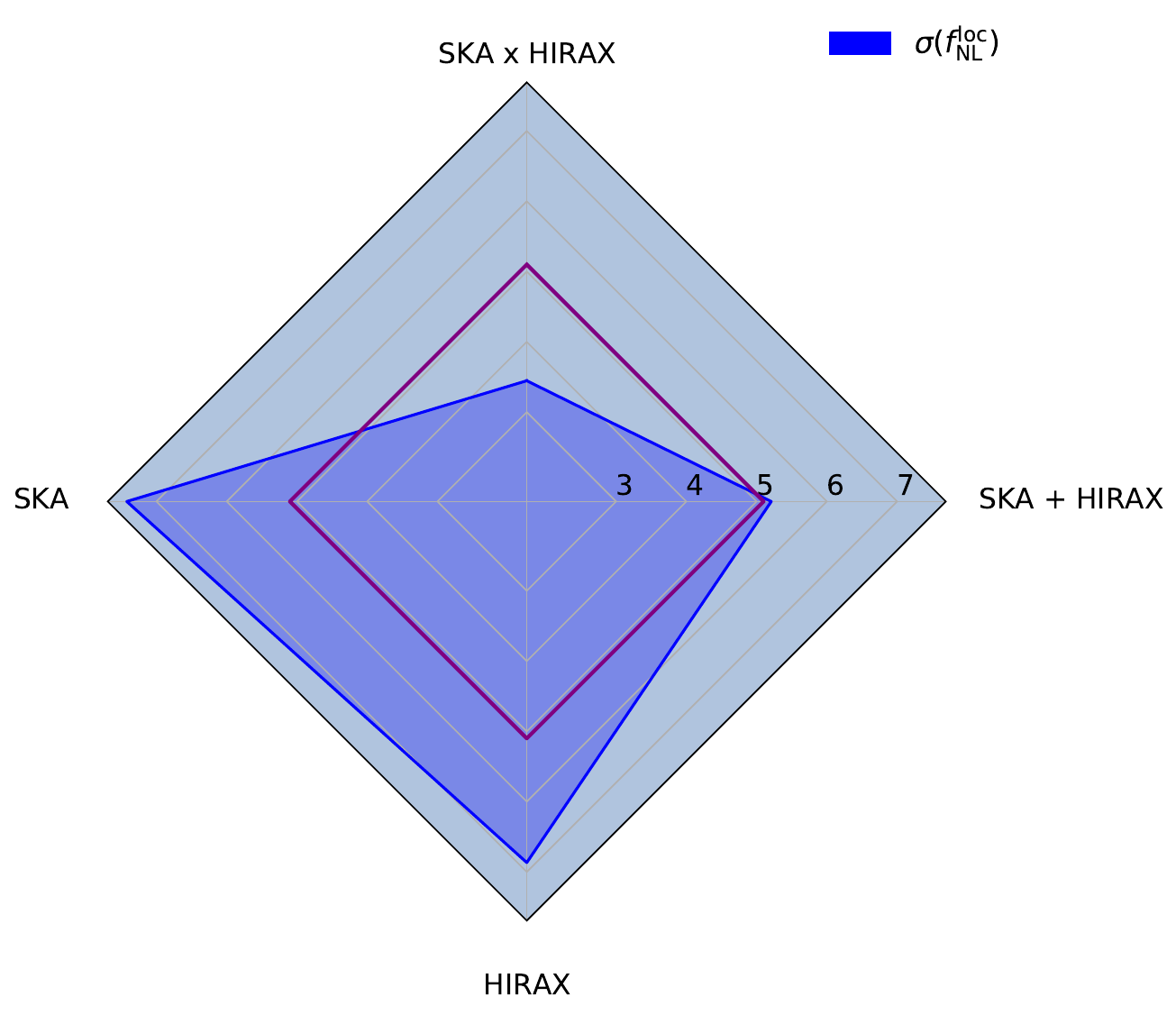}}
  \caption{Same as \Cref{fig:all_surveys_sigma_fnl}, but for all triangle shapes, not only squeezed, and with a fixed $k_{\rm max}(z)=0.2\;\Mpc$.}
\label{fig:all_surveys_sigma_fnl_all_triangles}
\end{figure*}

The advantages of the bimodal bispectrum are not limited to squeezed configurations and other shapes can be utilised. In \Cref{fig:all_surveys_sigma_fnl_all_triangles} we show forecasts on $\fnll$ when all triangle configurations are used and a redshift independent $k_{\rm max}=0.2\;\Mpc$ is considered. Note that the complete bimodal bispectrum covariance is used in these forecasts, since the approximate expression for the non-Gaussian part [\Cref{eq:Cov_NG}] is accurate, for the scales considered, in the case of all shapes of triangles \cite{Floss:2022wkq}. The bimodal bispectrum still provides a significant improvement over the constraints from each individual survey (a factor of $\sim 2$), while it exhibits a $\sim 30\%$ improvement over their summed signal. Pushing $k_{\rm max}$ to smaller values would enhance these differences even further, as shown in the main text.

\clearpage
\bibliographystyle{JHEP}
\bibliography{references}
\end{document}